\documentclass[aps,pra,reprint,numerical,amsmath,amssymb,amsfonts,showkeys,showpacs,preprintnumbers]{revtex4-1}
\usepackage{bm,graphicx,xcolor,hyperref,rotating}

\newcommand{\dpst}{\displaystyle}

\newcommand{\eps}{\varepsilon}
\newcommand{\om}{\omega}
\newcommand{\Ec}{E_{\rm c}}
\DeclareMathOperator{\Si}{Si}

\begin{document}

\title{A Tale of Two Electrons: Correlation at High Density}

\author{Pierre-Fran\c{c}ois Loos}
\email{loos@rsc.anu.edu.au}
\affiliation{Research School of Chemistry, 
Australian National University, Canberra, 
Australian Capital Territory, 0200, Australia}
\author{Peter M. W. Gill}
\thanks{Corresponding author}
\email{peter.gill@anu.edu.au}
\affiliation{Research School of Chemistry, 
Australian National University, Canberra, 
Australian Capital Territory, 0200, Australia}
\date{\today}

\begin{abstract}
We review our recent progress in the determination of the 
high-density correlation energy $\Ec$ in two-electron systems.
Several two-electron systems are considered, 
such as the well known helium-like ions (helium), 
and the Hooke's law atom (hookium).
We also present results regarding two electrons 
on the surface of a sphere (spherium), 
and two electrons trapped in a spherical box (ballium).
We also show that, 
in the large-dimension limit, 
the high-density correlation energy
of two opposite-spin electrons interacting 
{\em via} a Coulomb potential is given by 
$\Ec \sim -1/(8D^2)$ for any radial external 
potential $V(r)$, where $D$ is the dimensionality of the space.
This result explains the similarity of $\Ec$ in
the previous two-electron systems for $D=3$.
\end{abstract}


\maketitle

\section{
\label{sec:intro}
Introduction}

The Hartree-Fock (HF) approximation ignores the 
correlation between electrons, but gives roughly 99\%
of the total electronic energy \cite{Szabo}.
Moreover, it is often accurate for the prediction 
of molecular structure \cite{Helgaker},
computationally cheap and can be applied to large systems, 
especially within local (linear-scaling) strategies
\cite{White96,Schwegler96,Strain96,Burant96,Ochsenfeld98,
Kitaura99,Komeiji03,Fedorov06}.  To reduce the
computational cost still further, various numerical 
techniques have been developed including, for example, 
density fitting (or resolution of the identity)
\cite{Whitten73,Vahtras93,Dunlap77,Rendell94,Kendall97,Weigend02}, 
pseudospectral and Cholesky decomposition 
\cite{Martinez95,Friesner99,Beebe77,Roeggen86,Koch03,Aquilante07,Aquilante09},
dual basis methods \cite{Jurgens91,Wolinski03,Liang04,Steele06,Deng09,Deng10c},
and both attenuation \cite{CASE96, CAP96} and resolution \cite{Dombroski96, RO08,
Lag09, RRSE09} of the Coulomb operator.

Unfortunately, the part of the energy which the HF approximation ignores can have important chemical effects and this is particularly true when bonds are formed and/or broken.  Consequently, realistic model chemistries require a satisfactory treatment of electronic correlation.

The concept of electron correlation was introduced by Wigner \cite{Wigner34} and defined as
\begin{equation}
	\Ec = E - E_{\rm HF}
\end{equation}
by L\"owdin \cite{Lowdin59}, where $E$ is the exact non-relativistic energy.  Feynman refered to $\Ec$ as the ``stupidity energy'' \cite{Feynman72} because of the difficulties associated with its characterization in large systems.

Even though it is a formidable challenge to determine the correlation energy accurately, even in simple systems, recent heroic calculations on the helium atom \cite{Nakashima07, Nakashima08a, Nakashima08b, Kurokawa08} have demonstrated how near-exact results can be found.  Indeed, this elementary chemical system has been compared to the number $\pi$ by Charles Schwartz \cite{Schwartz06}:  {\em ``For thousands of years mathematicians have enjoyed competing with one other to compute ever more digits of the number $\pi$.  Among modern physicists, a close analogy is computation of the ground state energy of the helium atom, begun 75 years ago by E. A. Hylleraas.''}

Although $\Ec$ in the helium atom is now known very accurately, certain correlation effects remain incompletely understood and, for example, even the Coulomb hole \cite{Coulson61} itself is more subtle than one might imagine.  The primary effect of correlation is to decrease the likelihood of finding the two electrons close together and increase the probability of their being far apart.  However, accurate calculations have revealed the existence of a secondary Coulomb hole, implying that correlation also brings distant electrons closer together \cite{Pearson09a}.  The same observation has been made in the H$_2$ molecule by Per {\em et al.} \cite{Per09} and it appears that secondary (or long-range) Coulomb holes may be ubiquitous in two-electron systems \cite{TEOCS10}.

In order to get benchmark results for the development of intracule functional theory (IFT) \cite{Gill06, Dumont07, Crittenden07a, Crittenden07b, Bernard08, Pearson09b, Hollett10}, we have recently initiated an exhaustive study of two-electron systems \cite{BetheSalpeter}.  In the present Frontier Article, we review our recent progress in the determination of the correlation energy in various high-density two-electron systems:  the helium-like ions (Sec.~\ref{subsec:helium}), two electrons on the surface of a sphere (Sec.~\ref{subsec:spherium}), the Hooke's law atom (Sec.~\ref{subsec:hookium}), and two electrons trapped in a spherical box (Sec. \ref{subsec:ballium}).

It is reasonable to ask whether an understanding of the high-density regime is relevant to normal chemical systems but it turns out that most of the high-density behaviour of electrons is surprisingly similar to that at typical atomic and molecular electron densities.  Much can be learned about the languid waltz of a pair of electrons in a covalent bond from their frenetic jig in the high-density limit.  Moreover, it has led to an understanding of key systems, such as the uniform electron gas \cite{GellMann57,Vignale}, which form the cornerstone of the popular 
local density approximation in solid-state physics \cite{ParrYang}.

We also show (Sec.~\ref{sec:LDL}) that, in the large-dimension limit, the high-density correlation energy of two electrons is given by a simple universal rule which is independent of the external confining potential.  Just as one learns about interacting systems by studying non-interacting ones and then introducing the interaction perturbatively, one can understand our three-dimensional world by studying high-dimensional analogues and introducing dimension-reduction perturbatively. 

In this study, we confine our attention to the $^1S$ ground states of two-electron systems.  This allows us to ignore the spin coordinates and focus on the spatial part of the wave function.  Atomic units are used throughout.

\begin{figure}
\includegraphics[width=0.2\textwidth]{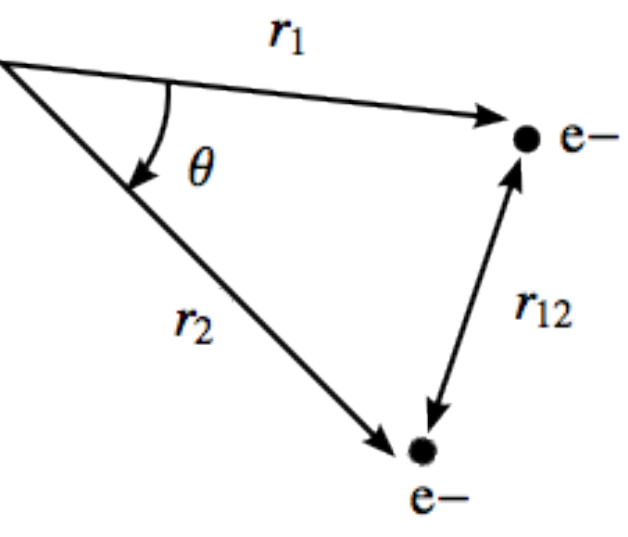}
\caption{
\label{fig:scheme}
The basic coordinates of a two-electron system.}
\end{figure}

\section{
\label{sec:HDL}
High-density limit}

For two electrons confined in a spherically-symmetric external potential $V(r)$, the Hamiltonian is
\begin{equation}
\label{H}
	\hat{H} =  
	- \frac{\nabla_1^2}{2} 
	- \frac{\nabla_2^2}{2} 
	+ V(r_1) + V(r_2) 
	+ \frac{1}{r_{12}},
\end{equation}
where the first two terms represent the kinetic energy of the electrons and $1/r_{12} = 1/\left| \bm{r}_1 - \bm{r}_2 \right|$ is the Coulomb operator (Fig.~\ref{fig:scheme}).  After a suitable scaling of the coordinates and energy \cite{EcLimit09, Proof}, the Hamiltonian can be recast as
\begin{equation}
\label{scaled-H}
	\hat{H} =  
	- \frac{\nabla_1^2}{2} 
	- \frac{\nabla_2^2}{2} 
	+ V(r_1) + V(r_2)
	+ \frac{1}{Z} \frac{1}{r_{12}},
\end{equation}
where $Z$ measures the confinement strength.  Equation \eqref{scaled-H} is well poised for a perturbation treatment in which the zeroth- and first-order Hamiltonians are 
\begin{align}
	\Hat{H}^{(0)} & = \Hat{h}_1^{(0)} + \Hat{h}_2^{(0)},&
	\Hat{H}^{(1)} & = \frac{1}{r_{12}},
\end{align}
and the one-electron Hamiltonian is given by
\begin{equation}
	\Hat{h}_i^{(0)} = - \frac{1}{2} \nabla_i^2 + V(r_i).
\end{equation}
The zeroth-order wave function satisfies the eigenequation
\begin{equation}
	\Hat{H}^{(0)} \Psi_0 \left(r_1,r_2\right) = E^{(0)} \Psi_0 \left(r_1,r_2\right),
\end{equation}
and the zeroth- and first-order energies are
\begin{gather}
	E^{(0)} = \left< \Psi_0 \left| \Hat{H}^{(0)} \right| \Psi_0 \right>,
	\label{E0}
	\\
	E^{(1)} = \left< \Psi_0 \left| \Hat{H}^{(1)} \right| \Psi_0 \right>.
	\label{E1}
\end{gather}
Following Hylleraas \cite{Hylleraas30}, we can use perturbation theory to expand both the exact \cite{Hylleraas30} and Hartree-Fock (HF) \cite{Linderberg61} energies as series in $1/Z$, yielding
\begin{multline} \label{Eex}
	E(Z,D,V) = E^{(0)}(D,V) Z^2 + E^{(1)}(D,V) Z	\\
				+ E^{(2)}(D,V) + E^{(3)}(D,V) Z^{-1} + \ldots,
\end{multline}
and
\begin{multline} \label{EHF}
	E_{\rm HF}(Z,D,V) = E_{\rm HF}^{(0)}(D,V) Z^2 + E_{\rm HF}^{(1)}(D,V) Z	\\
				+ E_{\rm HF}^{(2)}(D,V) + E_{\rm HF}^{(3)}(D,V) Z^{-1} + \ldots,
\end{multline}
where $D$ is the dimensionality of the space.  It is straightforward to show that
\begin{align}
	E^{(0)}(D,V) & = E_{\rm HF}^{(0)}(D,V),\\
	E^{(1)}(D,V) & = E_{\rm HF}^{(1)}(D,V)
\end{align}
and therefore, in the high-density (large-$Z$) limit, we find
\begin{align} \label{Ec}
	\Ec^{(2)}(D,V)	& = \lim_{Z\to\infty} \Ec(Z,D,V)								\notag	\\
			& = \lim_{Z\to\infty} \left[ E(Z,D,V)-E_{\rm HF}(Z,D,V) \right]	\notag	\\
			& = E^{(2)}(D,V) - E_{\rm HF}^{(2)}(D,V).
\end{align}

\begin{table*}
\caption{
\label{tab:E2}
Second-order energies and limiting correlation 
energies in two-electron systems.}
\begin{ruledtabular}
\begin{tabular}{lcccccc}
	System		&	$D=2$		&	$D=3$		&	$D=4$		&	$D=5$		&	$D=6$		&	$D=7$		\\
	\hline
			&			\multicolumn{6}{c}{Second-order exact energies, $-E^{(2)}(D,V)$, from \eqref{Eex}}					\\
	Helium		&	0.632740	&	0.157666 	&	0.070044	&	0.039395 	&	0.025208	&	0.017501	\\
	Spherium	&	0.227411 	&	0.047637 	&	0.019181	&	0.010139	&	0.006220	&	0.004189	\\
	Hookium		&	0.345655	&	0.077891 	&	0.032763	&	0.017821	&	0.011153	&	0.007622	\\
	Ballium		&	0.057959	&	0.014442	&	0.006194	&	0.003333	&	0.002037	&	0.001352	\\
	\hline
			&			\multicolumn{6}{c}{Second-order HF energies, $-E^{(2)}_{\rm HF}(D,V)$, from \eqref{EHF}}				\\
	Helium		&	0.412607	&	0.111003 	&	0.051111	&	0.029338	&	0.019020	&	0.013325	\\
	Spherium	&	0 		&	0 		&	0		&	0		&	0		&	0		\\
	Hookium		&	0.106014	&	0.028188 	&	0.012904	&	0.007382	&	0.004776	&	0.003342	\\
	Ballium		&	0.324120	&	0.069618	&	0.028107	&	0.014770	&	0.008977	&	0.005983 	\\
	\hline
			&			\multicolumn{6}{c}{Limiting correlation energies $-\Ec^{(2)}(D,V)$, from \eqref{Ec}}					\\
	Helium		&	0.220133	&	0.046663	&	0.018933	&	0.010057	&	0.006188	&	0.004176	\\
	Spherium	&	0.227411	&	0.047637	&	0.019181	&	0.010139	&	0.006220	&	0.004189	\\
	Hookium		&	0.239641	&	0.049703	&	0.019860	&	0.010439	&	0.006376	&	0.004280	\\
	Ballium		&	0.266161	&	0.055176	&	0.021913	&	0.011437	&	0.006940	&	0.004631	\\
\end{tabular}
\end{ruledtabular}
\end{table*}

\subsection{
\label{subsec:helium}
Helium}

As a first example, we consider the $D$-dimensional helium-like ions (He) \cite{Hylleraas30,Hylleraas64} where the electrons move in the Coulomb field of a nucleus with charge $Z$, {\em i.e.}
\begin{equation}
\label{V-He}
	V(r) = - \frac{Z}{r}.
\end{equation}
From the foregoing Section, we have
\begin{equation}
	\Hat{h}_0 = 
	- \frac{1}{2} 
	\left[ \frac{d^2}{dr^2} + \frac{D-1}{r} \frac{d}{dr} \right] - \frac{1}{r},
\end{equation}
and the zeroth-order wave function is
\begin{equation}
	\Psi_0(r_1,r_2) = \frac{4^D}{(D-1)^{D}\Gamma(D)} \exp \left[ - \frac{2(r_1+r_2)}{D-1}\right].
\end{equation}
The $E^{(0)}$ and $E^{(1)}$ values are given by
\begin{align}
	E^{(0)}(D,{\rm He})	
	& = - \frac{4}{(D-1)^2},	\\
	E^{(1)}(D,{\rm He})	
	& = \frac{4}{(D-1)^2} \frac{\Gamma\left(D+\frac{1}{2}\right) 
	\Gamma\left(\frac{D+1}{2}\right)}
	{\Gamma(D+1) \Gamma \left(\frac{D}{2}\right)},
\end{align}
where $\Gamma$ is the Gamma function \cite{NISTbook}.  

To compute the second-order energy $E^{(2)}$, we use the Hylleraas method \cite{Hylleraas30}, adopting the length and energy scaling of Herrick and Stillinger \cite{Herrick75} and employing the conventional Hylleraas basis functions \cite{Hylleraas30}
\begin{equation}
	\psi_{\om} \equiv \psi_{nlm} = s^n t^{2l} u^m \exp(-s/2),
\end{equation}
where $\om=(n,l,m)$ are non-negative integers and
\begin{align}
	s & = r_1 + r_2, &
	t & = r_1 - r_2,&
	u & = r_{12},
\end{align}
are the conventional Hylleraas coordinates.  The second-order energy, which minimizes the Hylleraas functional, is then given by
\begin{equation}
\label{E2-He}
	E^{(2)}(D,{\rm He}) = - \frac{1}{2} \mathbf{b^{\rm T}A^{-1}b},
\end{equation}
where
\begin{gather}
	\mathbf{A}_{\om_1 \om_2} = \mathbf{T}_{\om_1 \om_2} - \frac{D-1}{2} \mathbf{L}_{\om_1 \om_2}
				- 2 E^{(0)} \mathbf{S}_{\om_1 \om_2},
	\label{Hy-A}\\
	\mathbf{b}_{\om} = 2 E^{(1)} \mathbf{S}_{0 \om} - \frac{D-1}{2} \mathbf{U}_{0 \om},
	\label{Hy-B}
\end{gather}
In \eqref{Hy-A} and \eqref{Hy-B}, $\mathbf{T}$, $\mathbf{L}$, $\mathbf{S}$ and $\mathbf{U}$ are the kinetic, electron-nucleus, overlap and repulsion matrices, respectively, and are defined by
\begin{equation}
\begin{split}
	T_{\om_1\om_2} 	
	& = \frac{1}{2} \int \left[ \partial_s \psi_{\om_1}\partial_s \psi_{\om_2}  \right.
	+ \partial_t \psi_{\om_1}\partial_t \psi_{\om_2}  
	+ \partial_u \psi_{\om_1}\partial_u \psi_{\om_2}  \\
	& + s (u^2-t^2) \left( \partial_s \psi_{\om_1} \partial_u \psi_{\om_2} + \partial_u \psi_{\om_1} \partial_s \psi_{\om_2} \right)\\
	& \left. + t (s^2-t^2) \left( \partial_t \psi_{\om_1} \partial_u \psi_{\om_2} + \partial_u \psi_{\om_1} \partial_t \psi_{\om_2} \right)\right]d\tau
\end{split}
\end{equation}
and
\begin{gather}
	L_{\om_1\om_2} 	= 	\int \psi_{\om_1}\left( \frac{1}{r_1} + \frac{1}{r_2} \right)\psi_{\om_2} d\tau,		\\
	S_{\om_1\om_2} 	= 	\int  \psi_{\om_1}\psi_{\om_2} d\tau,								\\
	U_{\om_1\om_2} 	= 	\int \psi_{\om_1}\frac{1}{r_{12}}\psi_{\om_2} d\tau,
\end{gather}
with the volume element and domain of integration \cite{Herrick75}
\begin{gather}
	 d\tau	= 
			u \left(s^2-t^2\right) \mathcal{J}^{\frac{D-3}{2}} ds\,dt\,du,\\
	\mathcal{J} = \left(s^2-u^2\right) \left(u^2-t^2\right),\\
	\int d\tau = \int_0^{\infty} ds \int_0^{s} du \int_0^u dt.
\end{gather}

All the matrix elements can be obtained in closed form using the general formula
\begin{multline}
	\frac{4}{\Gamma(n + l + m + 2 D - 3)}
	\int s^n t^l u^m e^{-s} \mathcal{J}^{\frac{D-3}{2}} ds\,dt\,du	\\
	= 
	B\left(\frac{m+1}{2},\frac{D-1}{2}\right)
	B\left(\frac{l+m+D-1}{2},\frac{D-1}{2}\right),
\end{multline}
where
\begin{equation}
	B(m,n) = \frac{\Gamma(m)\Gamma(n)}{\Gamma(m+n)}
\end{equation}
is the beta function \cite{NISTbook}.  The $E_2$ value for $D=3$ has been studied in great detail \cite{Baker90,Kutzelnigg92}, but the only other helium whose $E_2$ value has been reported is 5-helium \cite{Herrick75} and this value was obtained by exploiting interdimensional degeneracies \cite{Herrick75b}.  Numerical values of $E^{(2)}$ for $2 \le D \le 7$ are given in Table \ref{tab:E2}.

$E^{(2)}_{\rm HF}$ values can be found by generalizing the Byers-Brown--Hirschfelder equations \cite{ByersBrown63} to obtain
\begin{gather}
	E^{(2)}_{\rm HF} 
	= - \int_0^\infty \frac{W(r)^2}{r^{D-1} \, \Psi_0(r,r)}\,dr,
	\label{E2HF-He}	\\
	W(r) = 2 \int_0^r [J(x) - E^{(1)}] \, \Psi_0(x,x) \, x^{D-1} \, dx,\\
	J(r) = \int_0^\infty \frac{\Psi_0(r,r)}{\max(r,x)} F\left[ \frac{3-D}{2},\frac{1}{2},\frac{D}{2},\alpha^2 \right] x^{D-1} dx.
	\label{J-He}
\end{gather}
where $\alpha = \frac{\min(x,r)}{\max(x,r)}$ and $F$ is the Gauss hypergeometric function \cite{NISTbook}.

$E_2^{\rm HF}(D,{\rm He})$ has been reported for $D=3$ by Linderberg \cite{Linderberg61} and Eq.~\eqref{E2HF-He} yields expressions such as
\begin{align}
	E^{(2)}_{\rm HF}(3,{\rm He})	& = \frac{9}{32} \ln \frac{3}{4} - \frac{13}{432},			\\
	E^{(2)}_{\rm HF}(5,{\rm He})	& = - \frac{903}{1024} \ln \frac{3}{4} - \frac{35\,213}{124\,416},	\\
	E^{(2)}_{\rm HF}(7,{\rm He})	& = - \frac{5\,643\,101}{204\,800} \ln \frac{3}{4} - \frac{640\,149\,405\,049}{80\,621\,568\,000}. 
\end{align}
Numerical values for $2 \le D \le 7$ are shown in Table \ref{tab:E2}.

\subsection{
\label{subsec:spherium}
Spherium}

Spherium (Sp) consists of two electrons, interacting {\em via} a Coulomb potential but constrained to remain on the surface of a sphere of radius $R=1/Z$ \cite{TEOAS09, Quasi09, Loos10, Excited10}.  This model was introduced by Berry and co-workers \cite{Ezra82, Ezra83, Ojha87, Hinde90} who used it to understand both weakly and strongly correlated systems, such as the ground and excited states of the helium atom, and also to suggest an alternating version of Hund's rule \cite{Warner85}.  Seidl studied this system in the context of density functional theory \cite{Seidl07b} to test the ISI (interaction-strength interpolation) model \cite{Seidl00}.  More recently, we have performed a comprehensive study of the spherium ground state, using electronic structure methods ranging from HF theory to explicitly correlated treatments \cite{TEOAS09}.

In this Section, we consider $D$-spherium, the generalization in which the two electrons are trapped on a $D$-sphere of radius $R$.  We adopt the convention that a $D$-sphere is the surface of a ($D+1$)-dimensional ball.  (Thus, for example, the Berry system is 2-spherium.)

Quantum mechanical models for which it is possible to solve exactly for a finite portion of the energy spectrum are said to be quasi-exactly solvable \cite{Ushveridze} and we have recently discovered that $D$-spherium is a member of this small but distinguished family \cite{Quasi09,Excited10}.  We have found that the Schr\"odinger equation for $D$-spherium can be solved exactly for a countably infinite set of $R$ values and that the resulting wave functions are polynomials in the interelectronic distance $r_{12} = |\bm{r}_1-\bm{r}_2|$.

The zeroth-order Hamiltonian of $D$-spherium is
\begin{equation}
	\Hat{H}_0 =  - \frac{d^2}{d\theta^2} - (D-1) \cot \theta \frac{d}{d\theta},
\end{equation}
where $\theta$ is the interelectronic angle and the associated eigenfunctions and eigenvalues are, respectively,
\begin{gather}
	\Psi_n(\theta) = \mathcal{N}\,C_n^{\frac{D-1}{2}}(\cos\theta),	\\
	\eps_n = n (n+D-1),
\end{gather}
where $C_n^{\frac{D-1}{2}}$ is a Gegenbauer polynomial \cite{NISTbook} and
\begin{equation}
	\mathcal{N} 
	= \sqrt{\frac{2^{D-3} (2n+D-1) \Gamma \left(\frac{D-1}{2}\right)^2 \Gamma (n+1)}{\pi \Gamma (n+D-1)}}.
\end{equation}
Using the partial-wave expansion of $r_{12}^{-1}$, one finds
\begin{equation}
	\left< C_0^{\frac{D-1}{2}} \Big| r_{12}^{-1} \Big| C_n^{\frac{D-1}{2}} \right> = \frac{(n+1)_{D-2}}{(n+\frac{1}{2})_{D-1}},
\end{equation}
where $(a)_n$ is the Pochhammer symbol \cite{NISTbook} 
\begin{equation}
	(a)_n = \frac{\Gamma(a+n)}{\Gamma(a)}
\end{equation}
From this, one can show that
\begin{align}
	E^{(0)}(D,{\rm Sp})	& = 0	\\
	E^{(1)}(D,{\rm Sp})	& = \frac{\Gamma(D-1) \Gamma\left(\frac{D+1}{2}\right)}
								{\Gamma\left(D-\frac{1}{2}\right) \Gamma\left(\frac{D}{2}\right)}
\end{align}
and the second-order energy is given by
\begin{equation} \label{E2-sp}
\begin{split}
	E^{(2)}(D,{\rm Sp})	
	& = \sum_{n=1}^{\infty} \frac{\left< \Psi_0 \left| r_{12}^{-1} \right| \Psi_{n}\right>^2}
	{\eps_0 - \eps_{n}}		\\
	& = - \frac{\Gamma(D)}{4\pi} \frac{\Gamma \left(\frac{D-1}{2}\right)^2}
	{\Gamma \left(\frac{D}{2}\right)^2}		\\
	& \times \sum_{n=1}^\infty \frac{(n+1)_{D-2}}{(n+\frac{1}{2})_{D-1}^2} 
	\left[\frac{1}{n}+\frac{1}{n+D-1}\right],
\end{split}
\end{equation}
which reduces to a generalized hypergeometric function.  It is also easy to show \cite{TEOAS09} that $E^{(2)}_{\rm HF}(D,{\rm Sp}) = 0$.

The $E^{(2)}$ (and thus $\Ec$) value for 2-spherium was first reported by Seidl \cite{Seidl07b} but elementary expressions for any $D$ can be obtained from Eq.~\eqref{E2-sp} and these are reported for $2 \le D \le 7$ in Table \ref{tab:E2-Sp}. 

\begin{table}
\caption{
\label{tab:E2-Sp}
$E^{(0)}$, $E^{(1)}$, $E^{(2)}$ and
$E_{\rm HF}^{(2)}$ for $D$-spherium.}
\begin{ruledtabular}
\begin{tabular}{ccccc}
$D$	&	$E^{(0)}$	&	$E^{(1)}$			&	$E^{(2)}$										&	$E_{\rm HF}^{(2)}$	\\[8pt]
\hline                                                                                                                                                                                           
2	&	0		&	1				&	$\dpst 4 \ln 2 - 3$									&	0			\\[8pt]
3	&	0		&	$\dpst \frac{8}{3\pi}$		&	$\dpst \frac{4}{3} - \frac{368}{27 \pi^2}$						&	0			\\[8pt]
4	&	0		&	$\dpst \frac{4}{5}$		&	$\dpst \frac{64}{75} \ln 2 - \frac{229}{375}$						&	0			\\[8pt]
5	&	0		&	$\dpst \frac{256}{105\pi}$	&	$\dpst \frac{24}{35} - \frac{2\,650\,112}{385\,875 \pi^2}$				&	0			\\[8pt]
6	&	0		&	$\dpst \frac{16}{21}$		&	$\dpst \frac{1024}{2205} \ln 2 - \frac{455\,803}{1\,389\,150}$				&	0			\\[8pt]
7	&	0		&	$\dpst \frac{8192}{3465\pi}$	&	$\dpst \frac{4924}{10\,395} - \frac{588\,637\,011\,968}{124\,804\,708\,875 \pi^2}$	&	0			\\[8pt]
\end{tabular}
\end{ruledtabular}
\end{table}

\begin{table*}
\caption{
\label{tab:E2-Ho}
$E^{(0)}$, $E^{(1)}$, $E^{(2)}$ and
$E_{\rm HF}^{(2)}$ for $D$-hookium.
($G$ is Catalan's constant \cite{NISTbook})}
\begin{ruledtabular}
\begin{tabular}{ccccc}
$D$	&	$E^{(0)}$	&	$E^{(1)}$		&	$E^{(2)}$		&	$E_{\rm HF}^{(2)}$							\\[4pt]
\hline                                                                                           
2	&	2		&	$\dpst \sqrt{\frac{\pi}{2}}$
				&	$\dpst 2G - \pi \ln 2$
				&	$\dpst -\frac{\pi}{32}{}_4F_3\left(1,1,\frac{3}{2},\frac{3}{2};2,2,2;\frac{1}{4}\right)$			\\[8pt]
3	&	3		&	$\dpst \sqrt{\frac{2}{\pi}}$			
				&	$\dpst 1 - \frac{2}{\pi}(1+\ln 2)$	
				&	$\dpst \frac{4}{3} - \frac{4}{\pi} \left[ 1 + \ln(8-4\sqrt{3}) \right]$								\\[8pt]
4	&	4		&	$\dpst \frac{1}{2}\sqrt{\frac{\pi}{2}}$		
				&	$\dpst \frac{4-\pi}{16} + \frac{2G - \pi \ln 2}{4}$
				&	$\dpst -\frac{\pi}{256}{}_4F_3\left(1,1,\frac{3}{2},\frac{3}{2};2,2,3;\frac{1}{4}\right)$			\\[8pt]
5	&	5		&	$\dpst \frac{2}{3}\sqrt{\frac{2}{\pi}}$		
				&	$\dpst \frac{5}{9} - \frac{8}{27\pi} (4+3 \ln 2)$
				&	$\dpst \frac{8}{27} - \frac{8}{27\pi} \left[ 8 - 3\sqrt{3} + 6 \ln(8-4\sqrt{3}) \right]$				\\[8pt]
6	&	6		&	$\dpst \frac{3}{8}\sqrt{\frac{\pi}{2}}$		
				&	$\dpst \frac{104-27\pi}{512} + \frac{9(2G-\pi\ln2)}{64}$
				&	$\dpst -\frac{3\pi}{2048}{}_4F_3\left(1,1,\frac{3}{2},\frac{3}{2};2,2,4;\frac{1}{4}\right)$		\\[8pt]
7	&	7		&	$\dpst \frac{8}{15}\sqrt{\frac{2}{\pi}}$	
				&	$\dpst \frac{89}{225} - \frac{128}{3375\pi} (23 + 15 \ln 2)$	
				&	$\dpst \frac{416}{675} - \frac{16}{3375\pi} \left[ 368 +15\sqrt{3} +240 \ln(8-4\sqrt{3}) \right]$	\\[8pt]
\end{tabular}
\end{ruledtabular}
\end{table*}

\subsection{
\label{subsec:hookium}
Hookium}

Hookium (Ho) consists of two electrons that repel Coulombically but are bound to the origin by the harmonic potential
\begin{equation}
\label{V-Ho}
	V(r) = \frac{Z^4}{2} r^2.
\end{equation}
This system was introduced 50 years ago by Kestner and Sinanoglu \cite{Kestner62} and solved analytically by Santos \cite{Santos68} and Kais {\em et al.} \cite{Kais89} for a particular value of the harmonic force constant.  Later, Taut showed that it is quasi-exactly solvable in that its Schr\"odinger equation can be solved for a countably infinite set of force constants \cite{Taut93}.  An interesting paper by Katriel {\em et al.} discusses similarities and differences between the hookium and helium atoms \cite{Katriel05}.

The one-electron Hamiltonian in $D$-hookium is
\begin{equation}
	\Hat{h}_0 
	= - \frac{1}{2} \left[ \frac{d^2}{dr^2} + \frac{D-1}{r} \frac{d}{dr} \right] 
	+ \frac{r^2}{2},
\end{equation}
and the zeroth-order wave functions are
\begin{equation}
	\Psi_\ell(r_1,r_2) = \prod_{k=1}^{D} \psi_{a_k} (x_{1,k}) \psi_{b_k} (x_{2,k}),
\end{equation}
where $x_{i,k}$ is the $k$th Cartesian coordinate of electron $i$, and $a_k$ and $b_k$ are non-negative integers.  The orbitals are the one-dimensional harmonic oscillator wave functions
\begin{equation}
	\psi_a(x) = \sqrt{2^a a! \pi^{1/2}} H_a(x) \exp(-x^2/2),
\end{equation}
where $H_a$ is the $a$th Hermite polynomial \cite{NISTbook}.  The energy differences between the eigenstates are given by
\begin{equation}
\label{excitation}
	\eps_\ell - \eps_0 = \sum_{k=1}^D (a_k+b_k) = 2n,
\end{equation}
where $2n$ is the excitation level, 
\textit{i.e.}~the number of nodes in $\Psi_\ell$.
It is not difficult to show that 
\begin{align}
	E^{(0)}(D,{\rm Ho}) & = D,\\
	E^{(1)}(D,{\rm Ho}) & = \frac{1}{\sqrt{2}} \frac{\Gamma(\frac{D-1}{2})}{\Gamma(\frac{D}{2})}.
\end{align}

Both $E^{(2)}$ and $E^{(2)}_{\rm HF}$ can be found by direct summation \cite{HookCorr05}, as in Eq.~\eqref{E2-sp}.  The sum includes all single and double excitations for $E^{(2)}$, but only singles for $E^{(2)}_{\rm HF}$.  The integral $\left< \Psi_0 \left| r_{12}^{-1} \right| \Psi_\ell \right>$ vanishes unless all of the $a_k+b_k$ are even and, in that case, it is given by
\begin{multline}
\label{int-ho}
	\left< \Psi_0 \left| r_{12}^{-1} \right| \Psi_\ell \right> =
	\frac{1}{\sqrt{2 \pi}} \frac{\Gamma \left(\frac{D-1}{2}\right) \Gamma \left(n+\frac{1}{2}\right) }{\Gamma(n+2)} \\
	\times \prod_{k=1}^D \frac{i^{a_k-b_k} }{\sqrt{\pi a_k! b_k!}} \Gamma \left(\frac{a_k+b_k+1}{2}\right).
\end{multline}
In this way, one eventually finds
\begin{gather}
	E^{(2)}(D,{\rm Ho})		
	= -\frac{\Gamma \left(\frac{D-1}{2}\right)^2}{4\,\Gamma\left(\frac{D}{2}\right)^2}
 	\sum_{n=1}^\infty \frac{\left(\frac{1}{2}\right)_n^2}{\left(\frac{D}{2}\right)_n} \frac{1}{n!\,n},
	\label{E2-ho}	\\
	E^{(2)}_{\rm HF}(D,{\rm Ho})	
	= -\frac{\Gamma \left(\frac{D-1}{2}\right)^2}{2\,\Gamma\left(\frac{D}{2}\right)^2}
	\sum_{n=1}^\infty \frac{\left(\frac{1}{2}\right)_n^2}{\left(\frac{D}{2}\right)_n} \frac{(1/4)^n}{n!\,n},
	\label{E2HF-ho}
\end{gather}
which reduce to generalized hypergeometric functions.

$E^{(2)}(3,{\rm Ho})$ has been derived by several groups \cite{White70,Cioslowski00,HookCorr05}, and the energies for other $D$ have been reported in Ref. \cite{EcLimit09}.  Closed-form expressions for $E^{(2)}$ and $E_{\rm HF}^{(2)}$, for $2 \le D \le 7$, are listed in Table \ref{tab:E2-Ho}.

\subsection{
\label{subsec:ballium}
Ballium}

Ballium (Ba) was first studied by Alavi and co-workers \cite{Alavi00, Thompson02, Thompson05} and consists of two electrons, repelling Coulombically, but confined within a ball of radius $R=1/Z$.  It has been used for the assessment of density-functional approximations \cite{Thompson02, Jung03, Jung04} and the study of Wigner molecules \cite{Wigner34} at low densities \cite{Thompson04a, Thompson04b, Thompson05}.  We recently obtained near-exact energies for various values of $R$ \cite{Ball}.

The one-electron Hamiltonian for $D$-ballium is
\begin{equation}
\label{H0}
	\Hat{h}_0 = - \frac{1}{2} \left[ \frac{d^2}{dr^2} + \frac{D-1}{r} \frac{d}{dr} \right] + V(r),
\end{equation}
and the external potential is defined by
\begin{equation}
\label{V}
	V(r) =	\begin{cases}
		0,		&	\text{ if } r < R,	\\
		\infty,		&	\text{otherwise},
		\end{cases}
\end{equation}
or equivalently
\begin{equation}
\label{V-bis}
	V(r) = (r/R)^m, \quad m \to \infty.
\end{equation}
Any physically acceptable eigenfunction of \eqref{H} 
must satisfy the Dirichlet boundary condition 
\begin{equation}
\label{Dirichlet}
	\Psi(r_1=R) = \Psi(r_2=R) = 0.
\end{equation}
The associated zeroth-order wave function 
of the zeroth-order Hamiltonian \eqref{H0} is
\begin{equation} \label{Psi0-Ba}
	\Psi_0(r_1,r_2) = 
	\frac{2}{J_{D/2}(\kappa)^2} 
	\frac{J_{D/2-1}(\kappa r_1)}{r_1^{D/2-1}}
	\frac{J_{D/2-1}(\kappa r_2)}{r_2^{D/2-1}}.
\end{equation}
In \eqref{Psi0-Ba}, $\kappa = j_{D/2-1,1}$ and 
$j_{\mu,k}$ is the $k$-th zero of the Bessel function 
of the first kind $J_{D/2-1}$ \cite{NISTbook}. 
The $E^{(0)}$ values are easily obtained 
from the relation
\begin{equation}
\label{E0-Ba}
	E^{(0)}(D,{\rm Ba}) = \kappa^2.
\end{equation}
For odd D, $E^{(1)}$ can be found in closed form 
{\em via} Eq. \eqref{E1}.
For example, for $D=3$,
\begin{equation}
	E^{(1)}(3,{\rm Ba}) 
	= 2 \left[1 - \frac{\Si(2\pi)}{2\pi} + \frac{\Si(4\pi)}{4\pi}\right],
\end{equation}
where $\Si$ is the sine integral function \cite{NISTbook}.

Using the basis functions
\begin{equation}
\label{psi-nlm}
 	\psi_{nlm} = (1-x^2) (1-y^2) x^{2n} y^{2l} z^m,
\end{equation}
where
\begin{align}
	x	& = r_1 / R,	&	
	y	& = r_2 / R,	&	
	z	& = r_{12} / R,
\end{align}
and $n$, $l$ and $m$ are non-negative integers, one finds that the second-order energy $E^{(2)}$ is given by \eqref{E2-He} where
\begin{gather}
	\mathbf{A} = \mathbf{T} - E^{(0)} \mathbf{S},	\\
	\mathbf{b} = \mathbf{C}^{\rm T} \left[ E^{(1)} \mathbf{S} - \mathbf{U} \right].
\end{gather}
The vector $\mathbf{C}$ contains the coefficients of the zeroth-order wave function \eqref{Psi0-Ba} expanded in the basis set \eqref{psi-nlm}.

The integrals needed to compute the different matrix elements are of the form
\begin{equation}
\label{Int}
	\mathcal{I}_{nlm} = \int x^n y^l z^m d\tau,
\end{equation}
with the volume element
\begin{gather}	
	d\tau = x\,y\,z\,\mathcal{J}^\frac{D-3}{2}\,dx\,dy\,dz 	\label{dV},	\\
	\mathcal{J} = (x+y+z)(x-y+z)(x+y-z)(x-y-z),
\end{gather}
and domain of integration
\begin{equation}
\label{intdV}
	\int d\tau = \int_0^1 dx \int_0^1 dy \int_{|x-y|}^{x+y} dz.
\end{equation}
One eventually finds
\begin{equation}
	\mathcal{I}_{nlm} = \sqrt{\pi}\frac{\Gamma\left(\frac{D-1}{2}\right)}{\Gamma\left(\frac{D}{2}\right)} 
	\frac{R^{n+l+m+2D}}{n+l+m+2D} \left(I_n^m+I_l^m\right),
\end{equation}
and
\begin{equation}
	I_a^b = \frac{{}_3F_2\left(\frac{a+D}{2},-\frac{b}{2},-\frac{b+D-2}{2}; \frac{a+D+2}{2},\frac{D}{2};1\right)}{a+D}.
\end{equation}
$E_{\rm HF}^{(2)}$ values can be found using the Byers-Brown--Hirschfelder equations (see Sec. \ref{subsec:helium}) and numerical values of $E^{(2)}$ and $E_{\rm HF}^{(2)}$ are listed in Table \ref{tab:E2}.

\section{
\label{sec:LDL}
Large-dimension limit}
\subsection{
\label{subsec:conjecture}
The conjecture}

In the large-$D$ limit, the quantum world reduces to a simpler semi-classical one \cite{Yaffe82} and problems that defy solution in $D=3$ sometimes become exactly solvable.  In favorable cases, such solutions provide useful insight into the $D=3$ case and this strategy has been successfully applied in many fields of physics \cite{Witten80,Yaffe83}.

Following Herschbach and Goodson \cite{Herschbach86,Goodson87}, we expand both the exact and HF energies with respect to $D$.
Although various asymptotic expansions exist \cite{Doren86} for this dimensional expansion, it is convenient \cite{Yaffe83} to write \cite{Doren86,Doren87,Goodson92,Goodson93}
\begin{align}
	E^{(2)}(D,V)
	& = \frac{E^{(2,0)}(V)}{D^2}
	+ \frac{E^{(2,1)}(V)}{D^3}
	+ \ldots,
	\label{E2DV}
	\\
	E_{\rm HF}^{(2)}(D,V)
	& = \frac{E_{\rm HF}^{(2,0)}(V)}{D^2}
	+ \frac{E_{\rm HF}^{(2,1)}(V)}{D^3}
	+ \ldots,
	\label{EHF2DV}
	\\
	\Ec^{(2)}(D,V)
	& = \frac{\Ec^{(2,0)}(V)}{D^2}
	+ \frac{\Ec^{(2,1)}(V)}{D^3}
	+ \ldots,
	\label{Ec2DV}
\end{align}
where
\begin{align}
	\Ec^{(2,0)}(V) 
	& = E^{(2,0)}(V) - E_{\rm HF}^{(2,0)}(V),
	\\
	\Ec^{(2,1)}(V) 
	& = E^{(2,1)}(V) - E_{\rm HF}^{(2,1)}(V).
\end{align}
Such double expansions of the correlation energy were originally introduced for the helium-like ions, and have led to accurate estimations of correlation \cite{Loeser87a, Loeser87b} and atomic energies \cite{Loeser87c, Kais93} {\em via} interpolation and renormalization techniques.  Equation \eqref{Ec2DV} applies equally to the $^1S$ ground state of any two-electron system confined by a spherical potential $V(r)$.

For helium, it is known \cite{Mlodinow81, Herschbach86, Goodson87} that
\begin{align}
	\Ec^{(2,0)}({\rm He}) & = - \frac{1}{8},
	&
	\Ec^{(2,1)}({\rm He}) & = - \frac{163}{384},
\end{align}
and we have recently found \cite{EcLimit09} that
\begin{align}
	\Ec^{(2,0)}({\rm Ho}) & = - \frac{1}{8},
	&
	\Ec^{(2,1)}({\rm Ho}) & = - \frac{111}{256},\\
	\Ec^{(2,0)}({\rm Sp}) & = - \frac{1}{8},
	&
	\Ec^{(2,1)}({\rm Sp}) & = - \frac{53}{128},\\
	\Ec^{(2,0)}({\rm Ba}) & = - \frac{1}{8},
	&
	\Ec^{(2,1)}({\rm Ba}) & = - \frac{85}{128}.
\end{align}
The fact that $\Ec^{(2,0)}$ is invariant to the external potential and $\Ec^{(2,1)}$ depends only weakly on it explains why the high-density correlation energies (Table \ref{tab:E2}) of all the systems are similar, though not identical, for $D=3$ \cite{EcLimit09,Ball}.

On this basis, we conjectured \cite{EcLimit09} that
\begin{equation}
\label{conjecture}
        \Ec^{(2)}(D,V) 
	\sim - \frac{1}{8D^2} - \frac{C(V)}{D^3}
\end{equation}
holds for \emph{any} spherical confining potential, where the coefficient $C(V)$ varies slowly with $V(r)$.

\subsection{
\label{subsec:proof}
The proof}

In this Section, we will summarize our proof of the conjecture \eqref{conjecture}.  More details can be found in Ref. \cite{Proof}.  We prove that $\Ec^{(2,0)}$ is universal, and that, for large $D$, the high-density correlation energy of the $^1S$ ground state of two electrons 
is given by \eqref{conjecture} for any confining potential of the form
\begin{equation}
\label{V-proof}
	V(r) = \text{sgn}(m) r^m v(r),
\end{equation}
where $v(r)$ possesses a Maclaurin series expansion
\begin{equation}
	v(r)	= v_0 + v_1 r + v_2 \frac{r^2}{2} + \ldots.
\end{equation}

After transforming both the dependent and independent variables \cite{Proof}, the Schr\"odinger equation can be brought to the simple form
\begin{equation}
\label{Hersch-trans}
	\left( \frac{1}{\Lambda} \Hat{\mathcal{T}} 
	+ \Hat{\mathcal{U}} 
	+ \Hat{\mathcal{V}} 
	+ \frac{1}{Z} \Hat{\mathcal{W}} \right) \Phi_D
	= \mathcal{E}_D \Phi_D,
\end{equation}
in which, for $S$ states, the kinetic, centrifugal, external potential and Coulomb operators are, respectively,
\begin{gather}
	-2 \Hat{\mathcal{T}} = 
	\left( \frac{\partial^2}{\partial r_1^2} + \frac{\partial^2}{\partial r_2^2} \right)
	+ \left( \frac{1}{r_1^2} + \frac{1}{r_1^2} \right) 
	\left( \frac{\partial^2}{\partial \theta^2} + \frac{1}{4} \right),
	\\
	\Hat{\mathcal{U}} = 
	\frac{1}{2 \sin^2 \theta} 
	\left( \frac{1}{r_1^2} + \frac{1}{r_1^2} \right),
	\\
	\Hat{\mathcal{V}} =  
	V(r_1) + V(r_2),
	\\
	\Hat{\mathcal{W}} =
	\frac{1}{\sqrt{r_1^2 + r_2^2 -2 r_1 r_2 \cos \theta}},
\end{gather}
and the dimensional perturbation parameter is
\begin{equation}
	\Lambda = \frac{(D-2)(D-4)}{4}.
\end{equation}
In this form, double perturbation theory can be used to expand the energy in terms of both $1/Z$ and $1/\Lambda$.

For $D=\infty$, the kinetic term vanishes and the electrons settle into a fixed (``Lewis'') structure \cite{Herschbach86} that minimizes the effective potential
\begin{equation}
\label{X}
	\Hat{\mathcal{X}} 
	= \Hat{\mathcal{U}} + \Hat{\mathcal{V}} 
	+ \frac{1}{Z} \Hat{\mathcal{W}}.
\end{equation}
The minimization conditions are
\begin{gather}
	\frac{\partial \Hat{\mathcal{X}}(r_1,r_2,\theta)}{\partial r_1} =
	\frac{\partial \Hat{\mathcal{X}}(r_1,r_2,\theta)}{\partial r_2} = 0,
	\label{dW-r}
	\\
	\frac{\partial \Hat{\mathcal{X}}(r_1,r_2,\theta)}{\partial \theta} = 0,
	\label{dW-theta}
\end{gather}
and the stability condition implies $m > -2$.  Assuming that the two electrons are equivalent, the resulting exact energy is
\begin{equation}
	\label{Einf}
	\mathcal{E}_{\infty} 
	= \Hat{\mathcal{X}} (r_{\infty},r_{\infty},\theta_{\infty}).
\end{equation}
It is easy to show that
\begin{gather}
	r_{\infty} = \alpha + \frac{\alpha^2}{m+2} 
	\left(\frac{1}{2\sqrt{2}} - \Lambda \frac{m+1}{m} \frac{v_1}{v_0} \right) \frac{1}{Z}
	+ \ldots,
	\label{r-eq}
	\\
	\cos \theta_{\infty} = - \frac{\alpha}{4\sqrt{2}} \frac{1}{Z}
	+ \ldots,
	\label{tetha-eq}
\end{gather}
where $\alpha^{-(m+2)} = \text{sgn}(m) m v_0$.

For the HF treatment, we have $\theta_{\infty}^{\rm HF} = \pi/2$.  Indeed, the HF wave function itself is independent of $\theta$, and the only
$\theta$ dependence comes from the $D$-dimensional Jacobian, which becomes a Dirac delta function centered at $\pi/2$ as $D\to\infty$.  Solving \eqref{dW-r}, one finds that $r_{\infty}^{\rm HF}$ and $r_{\infty}$ are equal to second-order in $1/Z$.  Thus, in the large-$D$ limit, the HF energy is
\begin{equation}
	\label{EinfHF}
	\mathcal{E}_{\infty}^{\rm HF}
	= \Hat{\mathcal{X}} \left(r_{\infty}^{\rm HF},r_{\infty}^{\rm HF},\frac{\pi}{2}\right),
\end{equation}
and correlation effects originate entirely from the fact that $\theta_\infty$ is slightly greater than $\pi/2$ for finite $Z$.

\begin{table}
\caption{
\label{tab:Ec}
$E^{(2,0)}$, $E_{\rm HF}^{(2,0)}$, $\Ec^{(2,0)}$ 
and $\Ec^{(2,1)}$ coefficients for various systems
and $v(r) = 1$.}
\begin{ruledtabular}
\begin{tabular}{lrccccc}
System		&	$m$		&	$-E^{(2,0)}$		&	$-E_{\rm HF}^{(2,0)}$	&	$-\Ec^{(2,0)}$		&	$-\Ec^{(2,1)}$	\\
\hline                                                                          
Helium		&	$-1$		&		5/8		&	1/2			&	1/8			&	0.424479	\\
Airium		&	1		&		7/24		&	1/6			&	1/8			&	0.412767	\\
Hookium		&	2		&		1/4		&	1/8			&	1/8			&	0.433594	\\
Quartium	&	4		&		5/24		&	1/12			&	1/8			&	0.465028	\\
Sextium		&	6		&		3/16		&	1/16			&	1/8			&	0.486771	\\
Ballium		&	$\infty$	&		1/8		&	0			&	1/8			&	0.664063	\\
\end{tabular}
\end{ruledtabular}
\end{table}

\begin{figure}
\includegraphics[width=0.45\textwidth]{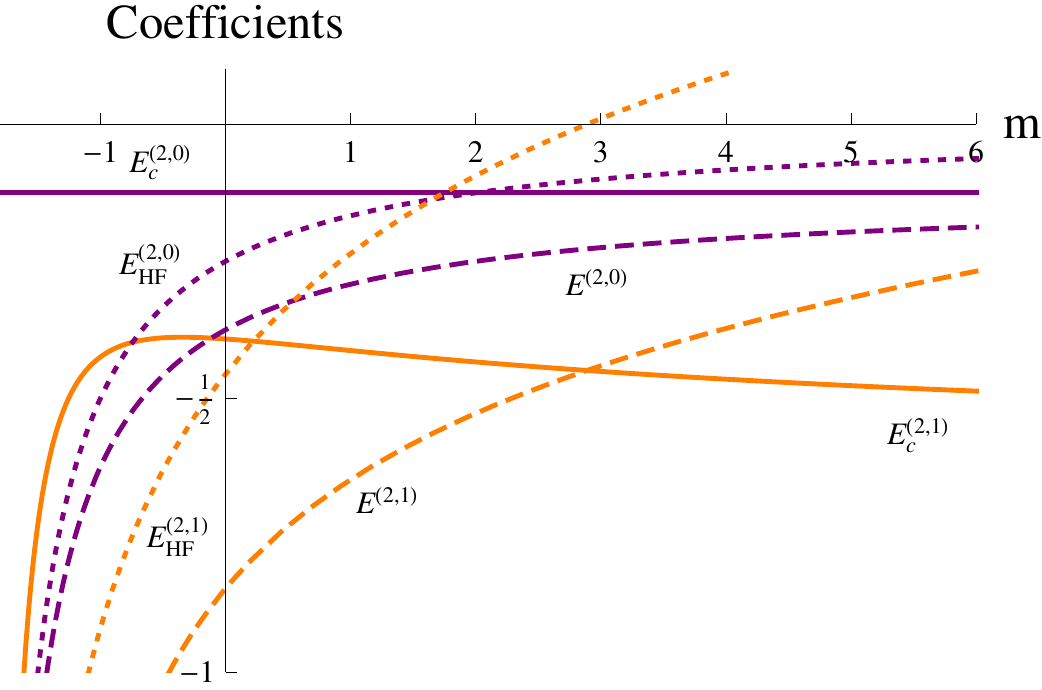}
\caption{\label{fig:Proof-fig} 
Coefficients of the exact (dashed), HF (dotted) and 
correlation (solid) energies with respect to $m$, 
for $v(r)=1$ (Eqs. \eqref{E2DV}, \eqref{EHF2DV} and \eqref{Ec2DV}).}
\end{figure}

Expanding \eqref{Einf} and \eqref{EinfHF} in terms of $Z$ and $D$ yields
\begin{align}
	E^{(2,0)}(V) 
	& = - \frac{1}{8} - \frac{1}{2(m+2)},
	\\
	E_{\rm HF}^{(2,0)}(V) 
	& = - \frac{1}{2(m+2)},
	\label{EHF20}
\end{align}
thus showing that both $E^{(2,0)}$ and $E_{\rm HF}^{(2,0)}$ depend on the leading power $m$ of the external potential but not on $v(r)$.

Subtracting these energies yields
\begin{equation}
	\Ec^{(2,0)}(V) = - \frac{1}{8} \label{Ec00},
\end{equation}
and this completes the proof that, in the high-density limit, the leading coefficient $\Ec^{(2,0)}$ of the large-$D$ expansion of the correlation energy is universal, {\em i.e.} it does not depend on the external potential $V(r)$.

The result \eqref{Ec00} is related to the cusp condition \cite{Kato57, Morgan93, Pan03}
\begin{equation}
\label{cusp}
	\left. \frac{\partial \Psi_D}{\partial r_{12}}\right|_{r_{12}=0} 
	= \frac{1}{D-1} \Psi_D(r_{12}=0),
\end{equation}
which arises from the cancellation of the Coulomb operator singularity by the $D$-dependent angular part of the kinetic operator \cite{Helgaker}.

The $E^{(2,1)}$ and $E_{\rm HF}^{(2,1)}$ coefficients can be found by considering the Langmuir vibrations of the electrons around their equilibrium positions \cite{Herschbach86,Goodson87}.  The general expressions depend on $v_0$ and $v_1$, but are not reported here.  However, for $v(r)=1$, which includes many of the most common external potentials, we find
\begin{multline}
	\Ec^{(2,1)}(V) = - \frac{85}{128} - \frac{9/32}{(m+2)^{3/2}} \\
	+ \frac{1/2}{(m+2)^{1/2}} + \frac{1/16}{(m+2)^{1/2}+2},
\end{multline}
showing that $\Ec^{(2,1)}$, unlike $\Ec^{(2,0)}$, is potential-dependent.  Numerical values of $\Ec^{(2,1)}$ are reported in Table \ref{tab:Ec} for various systems, and the components of the correlation energy are shown graphically in Fig. \ref{fig:Proof-fig}.

\section{
\label{sec:ccl}
Conclusion}

In this paper, we have reviewed our recent progress in the determination of the high-density correlation energy for four two-electron systems: the helium atom (He), the Hooke's law atom (Ho), two electrons confined on the surface of a sphere (Sp), and two electrons trapped in a ball (Ba).  In the large-$Z$ limit for $D=3$, we have found
\begin{equation}
	\Ec({\rm He}) \approx \Ec({\rm Sp}) \approx \Ec({\rm Ho}) \approx \Ec({\rm Ba}).
\end{equation}
These striking similarities can be rationalized by treating the dimensionality $D$ of space as a system parameter, and we have proved that, as $D$ grows, all such correlation energies exhibit the same universal behaviour
\begin{equation}
	\Ec \sim -1/(8D^2)
\end{equation}
in a $D$-dimensional space.  This is true {\em irrespective} of the nature of the external potential that confines the electrons.

\begin{acknowledgments}
We thank Yves Bernard, Joshua Hollett and Andrew Gilbert 
for kind support and many stimulating discussions.
P.M.W.G. thanks the NCI National Facility for a generous grant 
of supercomputer time and the Australian Research Council 
(Grants DP0771978 and DP0984806) for funding.
\end{acknowledgments}


\begin{thebibliography}{100}
\expandafter\ifx\csname url\endcsname\relax
  \def\url#1{\texttt{#1}}\fi
\expandafter\ifx\csname urlprefix\endcsname\relax\def\urlprefix{URL }\fi
\expandafter\ifx\csname href\endcsname\relax
  \def\href#1#2{#2} \def\path#1{#1}\fi

\bibitem{Szabo}
A.~Szabo, N.~S. Ostlund, Modern Quantum Chemistry : Introduction to Advanced
  Structure Theory, Dover publications Inc., Mineola, New-York, 1989.

\bibitem{Helgaker}
T.~Helgaker, P.~J{\o}rgensen, J.~Olsen, Molecular Electronic-Structure Theory,
  John Wiley \& Sons, Ltd., 2000.

\bibitem{White96}
C.~White, B.~G. Johnson, P.~M.~W. Gill, M.~Head-Gordon, Chem. Phys. Lett. 253
  (1996) 268.

\bibitem{Schwegler96}
E.~Schwegler, M.~Challacombe, J. Chem. Phys. 105 (1996) 2726.

\bibitem{Strain96}
M.~C. Strain, G.~E. Scuseria, M.~J. Frisch, Science 271 (1996) 51.

\bibitem{Burant96}
J.~C. Burant, G.~E. Scuseria, M.~J. Frisch, J. Chem. Phys. 105 (1996) 8969.

\bibitem{Ochsenfeld98}
C.~Ochsenfeld, C.~A. White, M.~Head-Gordon, J. Chem. Phys. 109 (1998) 1663.

\bibitem{Kitaura99}
K.~Kitaura, E.~Ikeo, T.~Asada, T.~Nakano, M.~Uebayasi, Chem. Phys. Lett. 313
  (1999) 701.

\bibitem{Komeiji03}
Y.~Komeiji, T.~Nakano, K.~Fukuzawa, Y.~Ueno, Y.~Inadomi, T.~Nemoto,
  M.~Uebaysai, D.~G. Fedorov, K.~Kitaura, Chem. Phys. Lett. 372 (2003) 342.

\bibitem{Fedorov06}
D.~G. Fedorov, K.~Kitaura, Chem. Phys. Lett. 433 (2006) 182.

\bibitem{Whitten73}
J.~L. Whitten, J. Chem. Phys. 58 (1973) 4496.

\bibitem{Vahtras93}
O.~Vahtras, J.~Alml{\"o}f, M.~W. Feyereisen, Chem. Phys. Lett. 213 (1993) 514.

\bibitem{Dunlap77}
B.~I. Dunlap, W.~D. Connolly, J.~R. Sabin, Int. J. Quantum Chem. Symp. 11
  (1977) 81.

\bibitem{Rendell94}
A.~P. Rendell, T.~J. Lee, J. Chem. Phys. 101 (1994) 400.

\bibitem{Kendall97}
R.~A. Kendall, H.~A. Fruchtl, Theor. Chem. Acc. 97 (1997) 158.

\bibitem{Weigend02}
F.~Weigend, Phys. Chem. Chem. Phys. 4 (2002) 4285.

\bibitem{Martinez95}
T.~J. Martinez, E.~A. Carter, Modern Electronic Structure Theory, Advanced
  Series in Physical Chemistry, World Scientific, Singapore, 1995, p. 1132.

\bibitem{Friesner99}
R.~A. Friesner, R.~B. Murphy, M.~D. Beachy, M.~N. Ringnalda, W.~T. Pollard,
  B.~D. Dunietz, Y.~Cao, J. Phys. Chem. A 103 (1999) 1913.

\bibitem{Beebe77}
N.~H.~F. Beebe, J.~Linderberg, Int. J. Quantum Chem. 12 (1977) 683.

\bibitem{Roeggen86}
I.~Roeggen, E.~Wisloff-Nilssen, Chem. Phys. Lett. 132 (1986) 154.

\bibitem{Koch03}
H.~Koch, A.~S. de~Meras, T.~B. Pedersen, J. Chem. Phys. 118 (2003) 9481.

\bibitem{Aquilante07}
F.~Aquilante, T.~B. Pedersen, R.~Lindh, J. Chem. Phys. 126 (2007) 194106.

\bibitem{Aquilante09}
F.~Aquilante, L.~Gagliardi, T.~B. Pedersen, R.~Lindh, J. Chem. Phys. 130 (2009)
  154107.

\bibitem{Jurgens91}
R.~Jurgens-Lutovsky, J.~Alml{\"o}f, Chem. Phys. Lett. 178 (1991) 451.

\bibitem{Wolinski03}
K.~Wolinski, P.~Pulay, J. Chem. Phys. 118 (2003) 9497.

\bibitem{Liang04}
W.~Z. Liang, M.~Head-Gordon, J. Phys. Chem. A 108 (2004) 3206.

\bibitem{Steele06}
R.~P. Steele, R.~A. DiStasio, Y.~Shao, J.~Kong, M.~Head-Gordon, J. Chem. Phys.
  125 (2006) 074108.

\bibitem{Deng09}
J.~Deng, A.~T.~B. Gilbert, P.~M.~W. Gill, J. Chem. Phys. 130 (2009) 231101.

\bibitem{Deng10c}
J.~Deng, A.~T.~B. Gilbert, P.~M.~W. Gill, J. Chem. Phys. 133 (2010) 044116.

\bibitem{CASE96}
R.~D. Adamson, J.~P. Dombroski, P.~M.~W. Gill, Chem. Phys. Lett. 254 (1996)
  329--336.

\bibitem{CAP96}
P.~M.~W. Gill, R.~D. Adamson, Chem. Phys. Lett. 261 (1996) 105--110.

\bibitem{Dombroski96}
J.~P. Dombroski, S.~W. Taylor, P.~M.~W. Gill, J. Phys. Chem. 100 (1996) 6272.

\bibitem{RO08}
S.~A. Varganov, A.~T.~B. Gilbert, E.~Deplazes, P.~M.~W. Gill, J. Chem. Phys.
  128 (2008) 201104.

\bibitem{Lag09}
P.~M.~W. Gill, A.~T.~B. Gilbert, Chem. Phys. 356 (2009) 86--90.

\bibitem{RRSE09}
T.~Limpanuparb, P.~M.~W. Gill, Phys. Chem. Chem. Phys. 11 (2009) 9176--9181.

\bibitem{Wigner34}
E.~Wigner, Phys. Rev. 46 (1934) 1002--1011.

\bibitem{Lowdin59}
P.-O. L{\"o}wdin, Adv. Chem. Phys. 2 (1959) 207--322.

\bibitem{Feynman72}
R.~P. Feynman, Statistical mechanics, Addison-Wesley, 1989.

\bibitem{Nakashima07}
H.~Nakashima, H.~Nakatsuji, J. Chem. Phys. 127 (2007) 224104.

\bibitem{Nakashima08a}
H.~Nakashima, Y.~Hijikata, H.~Nakatsuji, J. Chem. Phys. 128 (2008) 154107.

\bibitem{Nakashima08b}
H.~Nakashima, H.~Nakatsuji, J. Chem. Phys. 128 (2008) 154108.

\bibitem{Kurokawa08}
Y.~I. Kurokawa, H.~Nakashima, H.~Nakatsuji, Phys. Chem. Chem. Phys. 10 (2008)
  4486.

\bibitem{Schwartz06}
C.~Schwartz, Int. J. Mod. Phys. E 15 (2006) 877.

\bibitem{Coulson61}
C.~A. Coulson, A.~H. Neilson, Proc. Phys. Soc. (London) 78 (1961) 831.

\bibitem{Pearson09a}
J.~K. Pearson, P.~M.~W. Gill, J.~Ugalde, R.~J. Boyd, Mol. Phys. 07 (2009) 1089.

\bibitem{Per09}
M.~C. Per, S.~P. Russo, I.~K. Snook, J. Chem. Phys. 130 (2009) 134103.

\bibitem{TEOCS10}
P.-F. Loos, P.~M.~W. Gill, Phys. Rev. A 81 (2010) 052510.

\bibitem{Gill06}
P.~M.~W. Gill, D.~L. Crittenden, D.~P. O'Neill, N.~A. Besley, Phys. Chem. Chem.
  Phys. 8 (2006) 15.

\bibitem{Dumont07}
E.~E. Dumont, D.~L. Crittenden, P.~M.~W. Gill, Phys. Chem. Chem. Phys. 9 (2007)
  5340.

\bibitem{Crittenden07a}
D.~L. Crittenden, P.~M.~W. Gill, J. Chem. Phys. 127 (2007) 014101.

\bibitem{Crittenden07b}
D.~L. Crittenden, E.~E. Dumont, P.~M.~W. Gill, J. Chem. Phys. 127 (2007)
  141103.

\bibitem{Bernard08}
Y.~A. Bernard, D.~L. Crittenden, P.~M.~W. Gill, Phys. Chem. Chem. Phys. 10
  (2008) 3447.

\bibitem{Pearson09b}
J.~K. Pearson, D.~L. Crittenden, P.~M.~W. Gill, J. Chem. Phys. 130 (2009)
  164110.

\bibitem{Hollett10}
J.~W. Hollett, P.~M.~W. Gill, Phys. Chem. Chem. Phys. (2010) submitted.

\bibitem{BetheSalpeter}
H.~A. Bethe, E.~E. Salpeter, Quantum Mechanics of One- and Two-Electron Atoms,
  Dover Publications Inc., Mineola, New-York, 1977.

\bibitem{GellMann57}
M.~Gell-Mann, K.~A. Brueckner, Phys. Rev. 106 (1957) 364.

\bibitem{Vignale}
G.~F. Giuliani, G.~Vignale, Quantum theory of electron liquid, Cambridge
  University Press, Cambridge, 2005.

\bibitem{ParrYang}
R.~G. Parr, W.~Yang, Density Functional Theory for Atoms and Molecules, Oxford
  University Press, 1989.

\bibitem{EcLimit09}
P.-F. Loos, P.~M.~W. Gill, J. Chem. Phys. 131 (2009) 241101.

\bibitem{Proof}
P.-F. Loos, P.~M.~W. Gill, Phys. Rev. Lett. (2010) in press \href
  {http://arxiv.org/abs/arXiv:1005.0676v4} {\path{arXiv:1005.0676v4}}.

\bibitem{Hylleraas30}
E.~A. Hylleraas, Z. Phys. 65 (1930) 209.

\bibitem{Linderberg61}
J.~Linderberg, Phys. Rev. 121 (1961) 816.

\bibitem{Hylleraas64}
E.~A. Hylleraas, Adv. Quantum Chem. 1 (1964) 1.

\bibitem{NISTbook}
F.~W.~J. Olver, D.~W. Lozier, R.~F. Boisvert, C.~W. Clark (Eds.), NIST handbook
  of mathematical functions, Cambridge University Press, New York, 2010.

\bibitem{Herrick75}
D.~R. Herrick, F.~H. Stillinger, Phys. Rev. A 11 (1975) 42.

\bibitem{Baker90}
J.~D. Baker, D.~E. Freund, R.~N. Hill, J.~D. {Morgan III}, Phys. Rev. A 41
  (1990) 1247.

\bibitem{Kutzelnigg92}
W.~Kutzelnigg, J.~D. {Morgan III}, J. Chem. Phys. 96 (1992) 4484.

\bibitem{Herrick75b}
D.~R. Herrick, J. Math. Phys. 16 (1975) 281.

\bibitem{ByersBrown63}
W.~{Byers Brown}, J.~O. Hirschfelder, Proc. Natl. Acad. Sci. USA 50 (1963)
  399--406.

\bibitem{TEOAS09}
P.-F. Loos, P.~M.~W. Gill, Phys. Rev. A 79 (2009) 062517.

\bibitem{Quasi09}
P.-F. Loos, P.~M.~W. Gill, Phys. Rev. Lett. 103 (2009) 123008.

\bibitem{Loos10}
P.-F. Loos, Phys. Rev. A 81 (2010) 032510.

\bibitem{Excited10}
P.-F. Loos, P.~M.~W. Gill, Mol. Phys. (2010) in press \href
  {http://arxiv.org/abs/arXiv:1004.3641v2} {\path{arXiv:1004.3641v2}}.

\bibitem{Ezra82}
G.~S. Ezra, R.~S. Berry, Phys. Rev. A 25 (1982) 1513.

\bibitem{Ezra83}
G.~S. Ezra, R.~S. Berry, Phys. Rev. A 28 (1983) 1989.

\bibitem{Ojha87}
P.~C. Ojha, R.~S. Berry, Phys. Rev. A 36 (1987) 1575.

\bibitem{Hinde90}
R.~J. Hinde, R.~S. Berry, Phys. Rev. A 42 (1990) 2259.

\bibitem{Warner85}
J.~W. Warner, R.~S. Berry, Nature 313 (1985) 160.

\bibitem{Seidl07b}
M.~Seidl, Phys. Rev. A 75 (2007) 062506.

\bibitem{Seidl00}
M.~Seidl, J.~P. Perdew, S.~Kurth, Phys. Rev. Lett. 84 (2000) 5070.

\bibitem{Ushveridze}
A.~G. Ushveridze, Quasi-Exactly Solvable Models in Quantum Mechanics, Institute
  of Physics Publishing, 1994.

\bibitem{Kestner62}
N.~R. Kestner, O.~Sinanoglu, Phys. Rev. 128 (1962) 2687.

\bibitem{Santos68}
E.~Santos, Anal. R. Soc. Esp. Fis. Quim. 64 (1968) 177.

\bibitem{Kais89}
S.~Kais, D.~R. Herschbach, R.~D. Levine, J. Chem. Phys 91 (1989) 7791.

\bibitem{Taut93}
M.~Taut, Phys. Rev. A 48 (1993) 3561.

\bibitem{Katriel05}
J.~Katriel, S.~Roy, M.~Springborg, J. Chem. Phys. 123 (2005) 104104.

\bibitem{HookCorr05}
P.~M.~W. Gill, D.~P. O'Neill, J. Chem. Phys. 122 (2005) 094110.

\bibitem{White70}
R.~J. White, W.~{Byers Brown}, J. Chem. Phys. 53 (1970) 3869--3879.

\bibitem{Cioslowski00}
J.~Cioslowski, K.~Penal, J. Chem. Phys. 113 (2000) 8434.

\bibitem{Alavi00}
A.~Alavi, J. Chem. Phys. 113 (2000) 7735.

\bibitem{Thompson02}
D.~C. Thompson, A.~Alavi, Phys. Rev. B 66 (2002) 235118.

\bibitem{Thompson05}
D.~C. Thompson, A.~Alavi, J. Chem. Phys. 122 (2005) 124107.

\bibitem{Jung03}
J.~Jung, J.~E. Alvarellos, J. Chem. Phys. 118 (2003) 10825.

\bibitem{Jung04}
J.~Jung, P.~Garcia-Gonzalez, J.~E. Alvarellos, R.~W. Godby, Phys. Rev. A 69
  (2004) 052501.

\bibitem{Thompson04a}
D.~C. Thompson, A.~Alavi, Phys. Rev. B 69 (2004) 201302.

\bibitem{Thompson04b}
D.~C. Thompson, A.~Alavi, J. Phys.: Condens. Matter 16 (2004) 7979.

\bibitem{Ball}
P.-F. Loos, P.~M.~W. Gill, J. Chem. Phys. 132 (2010) 234111.

\bibitem{Yaffe82}
L.~G. Yaffe, Rev. Mod. Phys. 54 (1982) 407.

\bibitem{Witten80}
E.~Witten, Physics Today 33 (1980) 38.

\bibitem{Yaffe83}
L.~G. Yaffe, Physics Today 36 (1983) 50.

\bibitem{Herschbach86}
D.~R. Herschbach, J. Chem. Phys. 84 (1986) 838.

\bibitem{Goodson87}
D.~Z. Goodson, D.~R. Herschbach, J. Chem. Phys. 86 (1987) 4997.

\bibitem{Doren86}
D.~J. Doren, D.~R. Herschbach, Phys. Rev. A 34 (1986) 2654.

\bibitem{Doren87}
D.~J. Doren, D.~R. Herschbach, J. Chem. Phys. 87 (1987) 443.

\bibitem{Goodson92}
D.~Z. Goodson, M.~L\'opez-Cabrera, D.~R. Herschbach, J.~D. {Morgan III}, J.
  Chem. Phys 97 (1992) 8481.

\bibitem{Goodson93}
D.~Z. Goodson, M.~L\'opez-Cabrera, Low-D regime: the one-dimensional limit,
  Dimensional Scaling in Chemical Physics, Kluwer Academic Publishers,
  Dordrecht, 1993, p. 115.

\bibitem{Loeser87a}
J.~G. Loeser, D.~R. Herschbach, J. Chem. Phys. 86 (1987) 2114.

\bibitem{Loeser87b}
J.~G. Loeser, D.~R. Herschbach, J. Chem. Phys. 86 (1987) 3512.

\bibitem{Loeser87c}
J.~G. Loeser, J. Chem. Phys. 86 (1987) 5635.

\bibitem{Kais93}
S.~Kais, S.~M. Sung, D.~R. Herschbach, J. Chem. Phys 99 (1993) 5184.

\bibitem{Mlodinow81}
L.~D. Mlodinow, N.~Papanicolaou, Ann. Phys. 131 (1981) 1.

\bibitem{Kato57}
T.~Kato, Commun. Pure Appl. Math. 10 (1957) 151.

\bibitem{Morgan93}
J.~D. {Morgan III}, The dimensional dependence of rates of convergence of
  Rayleigh-Ritz variational calculations on atoms and molecules, Dimensional
  Scaling in Chemical Physics, Kluwer Academic Publishers, Dordrecht, 1993, p.
  336.

\bibitem{Pan03}
X.-Y. Pan, V.~Sahni, J. Chem. Phys. 119 (2003) 7083.

\end{thebibliography}
\end{document}